\documentclass[singlecolumn,prl,superscriptaddress]{revtex4}

\usepackage{soul}

\usepackage{setspace, graphicx, color, amssymb, amsmath, subfigure, verbatim, ulem}
\usepackage[all]{xy}
\usepackage[colorlinks]{hyperref}
\pdfoutput=1

    \def \T {{\mathbb{T}}}
    \def \e {{\varepsilon}}

    \def \o {{\theta}}

    \def \O {{\Theta}}

    \def \l {{\lambda}}
    \def \L {{\Lambda}}

    \def \n {{\eta}}

    \def \r{{\rho}}

\newcommand{\beq}{\begin{equation}}
\newcommand{\eeq}{\end{equation}}
\newcommand{\beqr}{\begin{eqnarray}}
\newcommand{\eeqr}{\end{eqnarray}}
\newcommand{\beqrn}{\begin{eqnarray*}}
\newcommand{\eeqrn}{\end{eqnarray*}}
\newcommand{\beqn}{\begin{equation*}}
\newcommand{\eeqn}{\end{equation*}}
\newcommand{\bei}{\begin{itemize}}
\newcommand{\beii}{\begin{itemize} \item}
\newcommand{\eei}{\end{itemize}}
\newcommand{\bmei}{\begin{itemize} \compactlist}
\newcommand{\emei}{\end{itemize}}
\newcommand{\ben}{\begin{enumerate}}
\newcommand{\een}{\end{enumerate}}
\newcommand{\bes}{\begin{small}}
\newcommand{\ees}{\end{small}}
\newcommand{\bec}{\begin{center}}
\newcommand{\eec}{\end{center}}

\newcommand{\Eawk}[1]{\textcolor{cyan}{[E: awkward/rephrase these words]}}

\newcounter{kfoo}

\begin{document}

%Abstract without citations
\begin{abstract}

Highly connected recurrent neural networks often produce chaotic dynamics, meaning their precise activity is sensitive to small perturbations.  What are the consequences for how such networks encode streams of temporal stimuli?  On the one hand, chaos is a strong source of randomness, suggesting that small changes in stimuli will be obscured by intrinsically generated variability.  On the other hand, recent work shows that the type of chaos that occurs in spiking networks can have a surprisingly low-dimensional structure, suggesting that there may be ``room" for fine stimulus features to be precisely resolved.  Here we show that strongly chaotic networks produce patterned spikes that reliably encode time-dependent stimuli:  using a decoder sensitive to spike times on timescales of 10's of ms,  one can easily distinguish responses to very similar inputs.  Moreover, recurrence serves to distribute signals throughout chaotic networks so that small groups of cells can encode substantial
information about signals arriving elsewhere.  A conclusion is that the presence of strong chaos in recurrent networks does not prohibit precise stimulus encoding. 
\end{abstract}

\title{Revisiting chaos in stimulus-driven spiking networks: signal encoding and discrimination}

\author{Guillaume Lajoie}
\affiliation{University of Washington Institute for Neuroengineering}
\affiliation{Formerly at Max Plank Institute for Dynamics and Self-Organization, Dept.~of Nonlinear Dynamics}
%\affiliation{Bernstein Center for Computational Neuroscience}

\author{Kevin K.~Lin}
\affiliation{University of Arizona, School of Mathematics}

\author{Jean-Philippe Thivierge}
\affiliation{University of Ottawa, School of Psychology}

\author{Eric Shea-Brown}
\affiliation{University of Washington, Dept.~of Applied Mathematics}
\affiliation{University of Washington, Dept.~of Physiology and Biophysics}

\date{\today}
\maketitle

%--------------------------------------------------------------------
\section{Introduction}

Highly recurrent connectivity occurs throughout the brain.  In the cortex, many circuits operate in a  ``balanced state''  in which every neuron receives a large number of excitatory (E) and inhibitory (I) inputs.  This creates synaptic currents that cancel on average but feature strong fluctuations (see, e.g., \cite{Sha+98,Vreeswijk:1998p14451,Renartetal10}), giving rise 
 to sustained irregular spiking~\cite{Softky:1993uj}.
Well-established results show that such strongly recurrent networks operating in a
  balanced regime can produce {\it chaotic}
 dynamics in a range of settings, from
abstracted firing rate models with random
connectivity~\cite{Sompolinsky:1988p534} to networks
of spiking units with excitatory and inhibitory cell
classes~\cite{Vreeswijk:1998p14451,Monteforte:2010p11768,LathamRNN00,Luccioli:2012p17675}.
Chaos implies that the network dynamics depend very
sensitively on network states, so that
tiny perturbations to initial conditions explode to large effects over time.
As a consequence, when the same stimulus is presented to a chaotic network multiple times, it will fail to generate reproducible responses.  How can stimuli be encoded in the variable spike trains that result (c.f.~\cite{Stanley:2013p1102,Faisal:2008p1531})?  A central issue is
the relationship between trial-to-trial variability
and input discriminability: since exactly the
same sensory input can elicit different neural
responses from one trial to the next (as can
  distinct inputs), how can one decide which stimulus
 is driving a network based on its
  response?

  %@@@@@@@@@@@@@@@@@@@@@@@@@@@
 \begin{figure*}%[h!]
%\begin{center}
\includegraphics{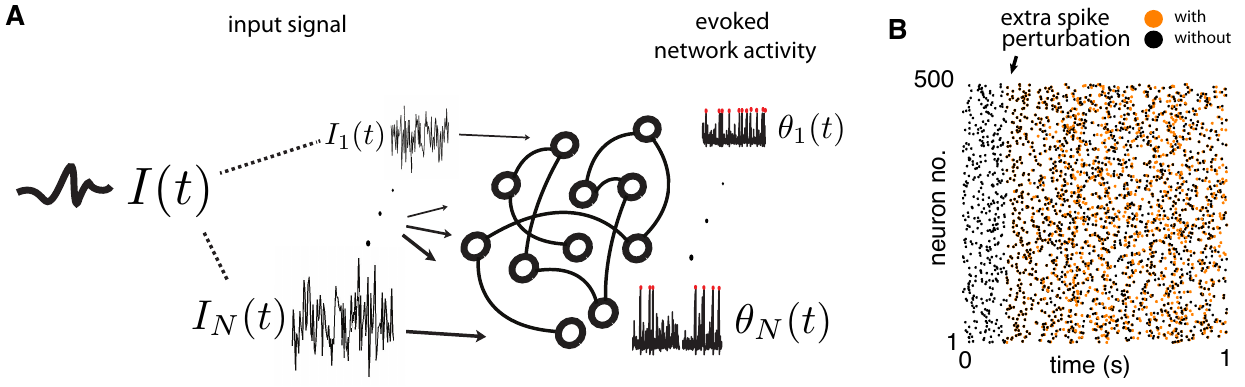}
\caption{({\bf A}) An input $I(t)$, with $N$ independent components, is
  presented to a chaotic network of $N$ spiking cells.  ({\bf B})  Raster plot of the network response to a fixed input $I(t)$, on two trials.  Arrow indicates a perturbation to the network on the second trial, in which an extra spike is ``added" to neuron number 1. Different
  color markers indicate the widely divergent spike rasters that occur with and
  without the perturbation.}
\label{fig:intro}
%\end{center}
\end{figure*}
%@@@@@@@@@@@@@@@@@@@@@@@@@@@
To illustrate the variable stimulus responses due to chaos, Figure~\ref{fig:intro} shows a model
balanced network (to be described in detail below)
driven by a fixed multi-dimensional stimulus, as well as
a raster plot of the spiking output of the network on
two trials. In each of these trials, the exact same
stimulus is presented, but on the second trial a small perturbation is
artificially introduced in the form of an extra spike
for neuron 1 (as in~\cite{Thivierge:2008p1353,London:2010p10818,Rosenbaum:2014p12}). This small perturbation quickly reverberates
throughout the whole network in a seemingly random
fashion~\cite{Thivierge:2008p1353,London:2010p10818,Rosenbaum:2014p12}.  If we were to repeat this process and
present a second input to this network, could a
decoder be trained to discriminate the spikes evoked
by the first input from the spikes evoked by the
second, despite both sets of output spikes being
subject to the type of variability shown here?  This is the central question that we
investigate here.

We study this stimulus coding question using a strongly chaotic, recurrent spiking network model driven by temporal stimuli. The strength of stimulus inputs is comparable to network interactions, so that dynamics are not dominated by external stimuli alone.
We find that, despite chaos, the network's spike patterns encodes temporal features of stimuli with sufficient precision so that the responses to close-by stimuli can  be accurately discriminated. We relate this coding precision to previous work grounded in the mathematical theory of dynamical systems, which shows that -- at the level of multi-neuron spike
patterns -- chaotic networks do not produce as much
variability as one might guess at first glance.  This is because such networks, when driven by time-dependent stimuli,
create low-dimensional chaotic attractors that show a
type of intermittent variability, with some spiking
``events'' predictably repeated across
trials~\cite{Lajoie:2013p17297,Lajoie:2014p18333}
(cf.~\cite{Marre:2009p16861,Rajan:2010p7924,PhysRevLett.69.3717}).
Our main findings are:
\begin{enumerate}
\item It is possible for strongly chaotic recurrent
  networks to produce multi-cell spike responses that remain
  discriminable even for highly similar inputs.
\item The same recurrent connections that produce chaos distribute stimulus information throughout the network, so that stimuli can be discriminated based on only a small subset of ``readout" neurons.
\item Input statistics influence the strength of chaotic fluctuations that can obscure stimuli.  We quantify this via a chaos-induced ``noise floor"; stimuli whose strength exceeds this floor will be easily discriminable.
\end{enumerate}

Our results are based on numerical simulations guided by mathematical theory, but connect to a broad experimental literature:  trial-to-trial variability in neural responses to repeated stimuli is often~\cite{Sha+98,Kara:2000p12837,Lu:2001p1087,Bair:2001p1086} (but not always~\cite{Lottem:2014p1082,Trussell:1999p1083,Scaglione:2011p1084,Tiesinga:2008p12846,Dan:1998p1085}) observed empirically.  Even though there are many likely
contributors to this variability, ranging from
stochasticity in neurotransmitter release or ion
channels (\cite{Faisal:2008p1563}, but
see~\cite{Bry+76,Mainen:1995p16589}) to confounding
factors like behavioral state, activity level, and
levels of adaptation~\cite{Niell:2008p1532,
  Mccormick:2015p1616}, chaotic interactions may play an inescapable role.  Indeed, chaos may represent a mechanism by which other sources of variability are amplified.

The remainder of this paper is organized as
  follows.  First, we briefly describe our network model and
the input discrimination task we use throughout (further details can be found in the {\it Methods} section). We then present an overview of dynamical
systems concepts useful to describe chaotic dynamics
and relate them to the spike outputs our model produces. Next, we describe the Tempotron classifier~\cite{Gutig:2006p10816}, which we train on the spikes produced by
our network. We use the performance of this classifier to quantify stimulus discrimination in chaotic networks, and how this depends on the temporal precision of the output spikes and the fraction of the network that is accessed by the decoder.  

%-------------------------------------------------------------------
\section{Results}

%----------
\subsection{Discrimination task and conceptual framework}

\subsubsection{Model overview}

We study a recurrent network of excitatory and inhibitory neurons with random, sparse coupling, as in~\cite{,Vreeswijk:1998p14451,Renartetal10,Monteforte:2010p11768,Lajoie:2013p17297}.
Every neuron $i=1,...,N$ in our
network receives and external input signal $I_i(t)$, which is a
time-dependent fluctuating stimulus that we describe
in more detail below. The collection of all these
signals is an $N$-dimensional input that we denote
$I(t)=\{I_i(t)\}_{i=1,...,N}$ and simply call the {\it
  network's input} or {\it stimulus}.  We emphasize that the inputs have $N$ independently varying components; i.e., they are $N$-dimensional. 
  
\begin{figure*}[h!]
%\begin{center}
\includegraphics{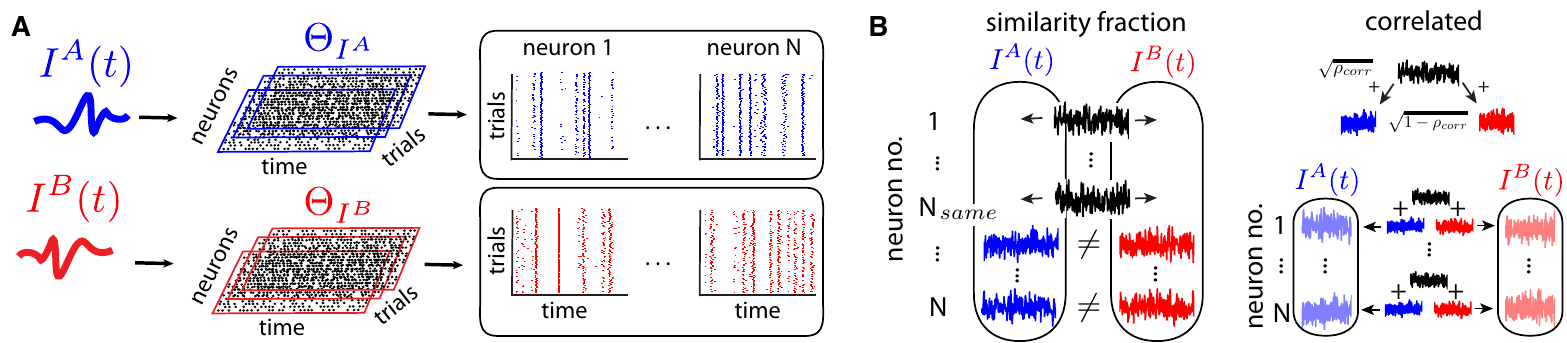}
\caption{({\bf A}) Two ensembles of spike patterns are generated by distinct, $N$-dimensional inputs $I^A$ and $I^B$ repeatedly presented to the network on multiple trials, where initial states are chosen at random. This results in two network-wide response ensembles $\O_{I^A}$ and $\O_{I^B}$ containing spike patterns across neurons, time and trials. ({\bf B}) Illustration of the two procedures used to control input similarity. Left: $\r_{same}$ controls the number of neurons $N_{same}=\r_{same}N$ that receive identical inputs $I_i(t)$ under both stimuli $I^A(t)$ and $I^B(t)$. Right: {\it correlation coefficient} $\r_{corr}$ controls the correlation of all pairs of neural inputs $I^A_i(t)$ and $I^B_i(t)$ (note that $I_i(t)$ and $I_j(t)$ remain uncorrelated if $i \neq j$).}
\label{fig:stims}
%\end{center}
\end{figure*}

 Our goal is to
explore the ability of a decoder to discriminate
between distinct network inputs, say $I^A(t)$ and
$I^B(t)$, based on the spike output from the whole network (or from a subset of cells from the network).  There are two main
factors that influence this discrimination: (i) the similarity of the stimuli $I^A(t)$ and $I^B(t)$, and (ii) the variability of the
network's response to a given stimulus from trial to trial.  Figure~~\ref{fig:stims} (A) illustrates this by showing the response of a few neurons in the
network to two distinct inputs, on several trials.

%@@@@@@@@@@@@@@@@@@@@@@@@@@@
%----------------------------
%@@@@@@@@@@@@@@@@@@@@@@@@@@@

%----------

Individual neurons are modelled as Quadratic-Integrate-and-Fire (QIF)
units~\cite{Ermentrout:1996p10447} which are separated into excitatory and
inhibitory populations (E and I) so that their
outgoing synaptic weights obey ``Dale's
Law"~\cite{kandel2013principles}.  We set 20$\%$ of the
neurons to be inhibitory, $80\%$ excitatory, and
couple the network according to the random and sparse, {\it
  balanced state}
architecture~\cite{vanVreeswijk:1996gja,Vreeswijk:1998p14451,Renartetal10,Monteforte:2010p11768,Lajoie:2013p17297}. Throughout
the paper, we report simulations carried out with
$N=500$ but note that the majority of the results we
outline are independent of network size, or scale with $N$ in simple ways. 
Network parameters are tuned so that the population-averaged mean firing rate of about $6.5$ Hz.
The detailed equations, parameter choices
and description of numerical methods used for
simulations can be found in the {\it Methods} section.

Briefly, the state of each neuron
$i=1,...,N$ is represented by a voltage variable $v_i
\in (-\infty,\infty)$.  These voltages evolve
according to intrinsic voltage-dependent (QIF) dynamics, smooth and localized synaptic
interactions, and external inputs $I_i(t)$.  For convenience and for technical reasons, we use
a smooth change of
coordinates~\cite{Ermentrout:1996p10447} that maps
these unbounded values to phase variables $\o_i \in
[0,1]$. Here $\o_i=0$ and $\o_i=1$ represent the same
state of $v_i = \pm \infty$, and a neuron is said to ``spike" once it reaches
this point. Mathematically, this means that the
voltage of each neuron is represented by a point on a
circle and the state of the entire network at time $t$
is given by the vector of phases
$\o(t)=(\o_1(t),...,\o_N(t))$. Thus, the state of all neurons in the network can be thought of as a point on an $N$-torus ($\T^N$). 
%----------
\subsubsection{Stimulus statistics and discrimination task}

We model the stimulus components $I_i(t)$ to individual
neurons $i$ as a realization of Gaussian white noise,
which is fixed and exactly repeated on every trial,
i.e., the noise is ``frozen''. We emphasize
that even though input components $I_i(t)$'s are
modelled as random processes, they represent signals
and not noise.  In fact, there is no noise 
in our model:  it is a deterministic but non-autonomous dynamical system.  All trial-to-trial variability is
generated by chaos, as we explain in the next section.
Throughout, we require that the realizations $I_i(t)$
be statistically independent from each other across
neurons $i$, so that the network's input $I(t)=\{I_i(t)\}_i$ is
free of redundancies and can be thought of as a
$N$-dimensional signal.  The mean $\n$ and increment
variance $\e$ of each $I_i(t)$ are global parameters
that are fixed across all neurons.  They control the
statistics of the input by setting its DC component
and fluctuation amplitude, which both modulate the mean firing rate of the network.
Importantly, when we compare the network's response to two stimuli
$I^A(t)$ and $I^B(t)$, we also require $\e$ and $\n$
to be the same.  This way, the averaged statistics of network responses are identical for any stimulus pair, and discrimination must rely on the differences in random fluctuations between the two specific realizations $I^A(t)$ and $I^B(t)$.
Unless otherwise noted, we set the parameters to
$\n=-0.5$ and $\e=0.5$ (see {\it Methods} for more details).  We revisit their influence in
the section {\it Signal strength modulates a noise floor from chaos} below.

\smallskip
To study our network's sensitivity to changes in its
inputs, we introduce two notions of the similarity, or
``distance", between the stimuli $I^A(t)$ and $I^B(t)$. First, we define $\r_{same}\in[0,1]$ as the
proportion of the network's neurons that receive
identical inputs ($I^A_i(t)=I^B_i(t)$) under both $A$
and $B$ paradigms; the remaining fraction $1-\r_{same}$ receive
non-identical, independent inputs
($I^A_i(t)\neq I^B_i(t)$).  
Second, we vary the
correlation $\r_{corr}\in[0,1]$ between input pairs driving each neuron $i$, simultaneously for all $i$.
Thus, $I^A_i(t)$ and $I^B_i(t)$ are
jointly gaussian processes, each a realization of
white noise, with correlation coefficient equal to
$\r_{corr}$. Note that the inputs to distinct
cells remain independent regardless of stimulus choice
($\langle I^{A,B}_i(t)I^{A,B}_j(t)\rangle_t=0$ for
$i\neq j$) and that $\r_{corr}$ and $\r_{same}$ only control the
similarity of inputs to the same neuron across different stimuli. In both cases,
$\r_{corr}=\r_{same}=0$ enforces that all pairs
$(I_i^A(t),I_i^B(t))$ are independent, whereas if
$\r_{corr}=\r_{same}=1$, they are identical. Figure~\ref{fig:stims} (B)
illustrates the effect of both these similarity
parameters. 

\smallskip
As will be explained in detail, we evaluate the discriminability of our network's response using a linear classifier called the Tempotron~\cite{Gutig:2006p10816}. It can be interpreted as a simple integrating neuron, reading out spikes from the network, whose task is to fire a spike if stimulus $I^A(t)$ was presented and stay silent for $I^B(t)$. This classifier will be trained on ensembles of responses to both stimuli, and tested on novel responses. Using this framework, we also vary the number of readout neurons that the classifier has access to (i.e., the readout dimension). It is the relationship between the input statistics, similarity and readout dimension we seek to better understand.

\subsubsection{Chaos: a state space view of variability}
  
Suppose that an input $I(t)$ is presented to the
network starting at time $t=0$, but that the state of
the network at that time (i.e., the ``voltages" $\o^0=(\o_1^0,...,\o_N^0)$) is unknown.  What
can be said about the network's response to the ongoing input $I(t)$ for $t>0$? 
To approach this question, consider an ensemble of many initial network states $\o^0$. To this ensemble corresponds another one, consisting of network responses, i.e., a version of the network's activity parametrized by $t>0$ for each choice of initial state $\o^0$, in response to the same $I(t)$. Each ``response" in this ensemble represents a different ``trial," much like in an experiment where exactly the same stimulus is repeatedly presented to a system. Trial-to-trial variability thus depends on how distributed these ensembles of responses are about the network's state space.
    The theory of Random
    Dynamical Systems (RDS), which we briefly review
    below, can help us understand response ensembles.

\smallskip
Formally, we define a {\it response ensemble} associated with an input $I(t)$ as a collection of network trajectories through state space, for which initial states were sampled independently from some initial probability distribution. In this paper, we choose this distribution to be the uniform one, meaning that each point in state space has equal probability of being picked. We call individual trajectories within a response ensemble {\it trials}. Snapshots of all trials at any time $t>0$ are ensembles of points, corresponding to a probability
distribution that describes all possible network states given the system
has been subjected to the stimulus $I(t)$ up to time
$t$. Taken together for all $t>0$, these ``snapshot distributions"~\cite{Namenson:1996ck} define input- and time-dependent probability distributions describing the network's evoked activity for all possible initial states at once and they dictate statistical attributes of our network such as trial-to-trial variability.

We denote the response ensemble at time $t$, subject
to input $I(t)$, by $\O_I^t$. The variability of a
network's response to $I(t)$ is characterized by the
breadth of differences between the trajectories forming
its response ensemble, i.e., the ``size'' of $\O_I^t$ as measured by, e.g., its
  diameter. Although difficult to describe
by simple mathematical formulae, for
  specific systems, response ensembles are
well-defined mathematical objects with well-understood
  geometric properties that can be numerically
  characterized. RDS theory provides a
framework for studying these properties.  One of the
key results of RDS theory is that under very general
conditions, the ensembles $\O_I^t$ are concentrated on
time-evolving geometric objects known as {\it random
  attractors}, so called because almost all initial
conditions, when subjected to the same forcing $I(t)$,
will converge to the attractor.
Unlike ``classical attractors" in un-driven systems,
which have fixed positions in space (e.g. fixed points, periodic orbits, strange
attractors), the position of $\O_I^t$ changes in time
$t$ and with input choice $I(t)$. This is because our networks are driven by time-dependent
  stimuli, and are governed by non-autonomous systems
  of differential equations (see {\it Methods}).

Geometrically, these random attractors can be
  points, curves, or higher-dimensional surfaces, and
  can wind around state space in complicated ways.
  The exact position and shape of an attractor depends
  on both the system parameters and the specific
  realization of the stimulus. However, RDS
theory also states that certain important properties of
  the attractor ---and thus the response ensembles $\O_I^t$--- are
independent of specific choices of input $I(t)$,
are invariant in time, and depend only on system parameters.  An important one is the
sensitivity of network responses to small
  perturbations, as measured by the {\it Lyapunov
  exponents} of the system.  For an $N$-dimensional
system, these exponents are real numbers
$\lambda_1\geq\cdots\geq\lambda_N$ that describe the
rate of separation of nearby trajectories in different
state space directions.  For a system like our
network, the Lyapunov exponents do not depend on the
choice of input realization $I(t)$ so long as it is a
realization of white noise with same parameters $\e$
and $\n$.  A criterion for chaos is the presence of
positive Lyapunov exponents.  Moreover, the number of
positive Lyapunov exponents roughly indicates the
number of unstable directions in state space, and
their magnitude indicates how strong the amplification of
small perturbations is in those directions.  In a chaotic
system, almost any nearby trajectories will diverge
from each other exponentially fast, but they do so
only along unstable directions of attractors.

Other geometric properties of random attractors
  can be related to their Lyapunov exponents as well.
  For example, the {\it attractor dimension} is a
quantity describing how much of state space is occupied
by the attractor --- the source of trial-to-trial
variability for our network --- and is 
(roughly) given by the number of positive
Lyapunov exponents. To see that this should be
  the case, imagine a cloud of initial conditions
  evolving according to the same stimulus.  The state-space
  expansion associated with positive exponents tends
  to ``stretch'' this cloud, leading to the formation
  of smooth $\L^+$-dimensional surfaces, where $\L^+$ is the number of positive
  exponents (see, e.g., \cite{Lin:2013p1314} for a general,
non-technical introduction and
\cite{kifer,Arnold:2003p98} for more details).
If all exponents are negative, for example, then
  the attractor is just a (time-dependent) point,
  whereas the presence of a single positive exponent
  suggests that the attractor is curve-like, etc.

In previous work, we showed that in a wide variety of
dynamical regimes, the number of positive Lyapunov
exponents in driven balanced networks is less than
$20\%$ of the network's dimension $N$, and often
remains below $10\%$~\cite{
  Lajoie:2013p17297,Lajoie:2014p18333}.
In most of the paper below, we choose network and
input parameters, described above and in {\it Methods}, so
that about $8\%$ of Lyapunov exponents are
positive. Here, the rate of trajectory separation is strong with
$\l_1\simeq3.5$, and the network is by all accounts in
a chaotic regime.  At the same time, there are plenty of directions
in which the attractor is ``thin", leading to
trial-to-trial variability that is far from
homogeneously random at the population level. We
explore different parameter regimes, leading to
distinct attractor dimensions, toward the end of this
article.

%-------
\subsection{Response ensembles and state-space separability}

Before stating our main findings, we
  first present some numerical results characterizing
  the chaotic response ensembles for our model system, i.e., the ensemble of trajectories evoked by the same input but starting from different initial states.
  Our main goal will be to compute the typical
  ``diameter'' of a response ensemble, and to compare
  this statistic to the typical distance between
  response ensembles elicited by distinct stimuli.
From the discussion above, we know that for two inputs
$I^A(t)$ and $I^B(t)$ with identical statistics, the
dimension of their associated response ensembles, and thus their level of
trial-to-trial variability, will be the same.
Revisiting our discrimination task, network responses will be
discriminable if and only if the corresponding
ensembles $\O^t_{I^A}$ and $\O^t_{I^B}$ do not overlap
most of the time.  To predict when this
is the case, knowing only the dimension of $\O^t_{I^A}$ and $\O^t_{I^B}$ is
insufficient; we need to understand how the position
of $\O^t_I$ in state space depends on the choice of
$I(t)$. We will present evidence
  that for the balanced spiking networks studied here, there exist broad parameter
  regimes where the relation
\begin{displaymath}
  \mbox{{\it diameter of a response ensemble}}\ll
  \mbox{{\it distance between different ensembles}}
\end{displaymath}
generally holds.
\smallskip

\begin{figure}[h!]
%\begin{center}
\includegraphics{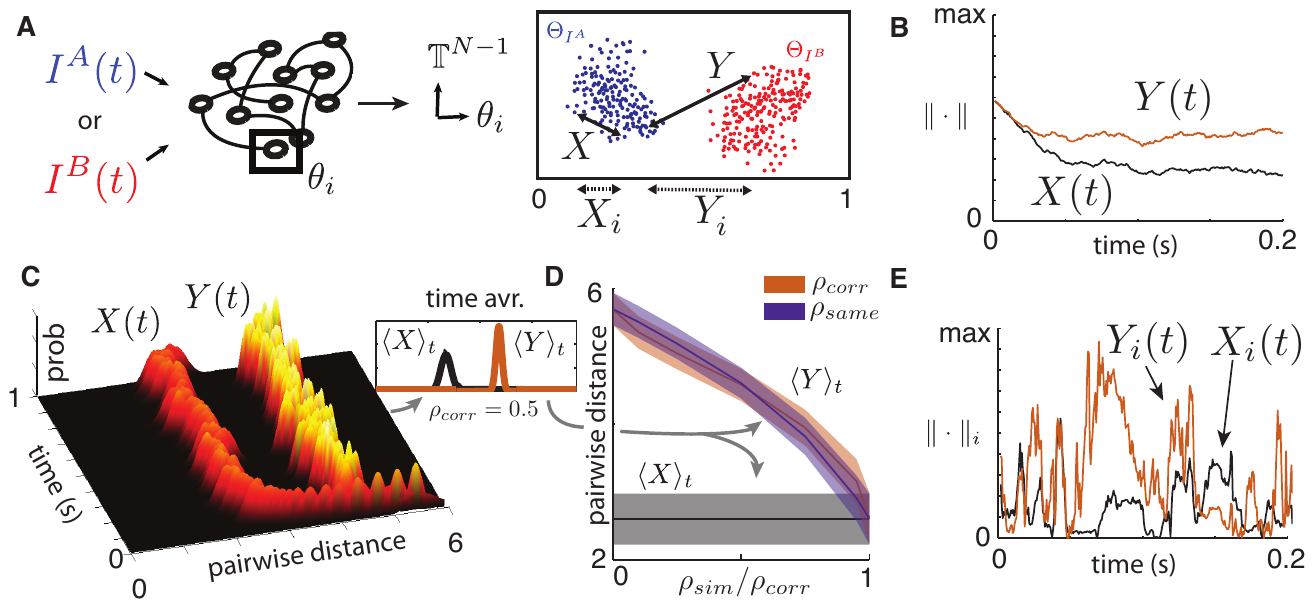}
\caption{({\bf A}) Illustration of pairwise distances
  between trajectories within an ensemble ($X$) and
  across ensembles ($Y$). Also illustrated are equivalent
  distances for a single neuron's state variable $\o_i$
  ($X_i$, $Y_i$). ({\bf B}) Pairwise distances between
  two trajectories for the entire network
  (on the $N$-dimensional torus $\T^N$, $N=500$) as a
  function of time.  ({\bf C}) Time evolution of
  pairwise distance distributions for entire network
  ($N=500$), sampled with $100$ trials from each
  ensemble. After a very fast transient ($\sim$ 10-50
  ms), distributions settle into stable values. For
  panels (B,C,D), $\rho_{corr}=0.5$. ({\bf D}) Mean of
  stationary pairwise distances for a range of
  similarity parameters $\r_{corr}$ and
  $\r_{same}$. Shaded areas show one standard
    deviation. Both mean and standard deviation
  were averaged over $1000$ time points between
  $t=100$ and $t=12,000$ ms.  ({\bf E}) Pairwise
  distances between two trajectories in a single
  neuron's coordinate $\o_i$ as a function of time.}
\label{fig:dist}
%\end{center}
\end{figure}

Consider the following pairwise distance quantities, the statistics of which we sample using numerically simulated network trajectories (see {\it Methods} for details):
\beq
\label{dist_RV}
\begin{split}
X(t)=&\|\o(t,\o_0,I^{A})-\o(t,\tilde{\o}_0,I^{A})\|\\
Y(t)=&\|\o(t,\o_0,I^{A})-\o(t,\tilde{\o}_0,I^{B})\|
\end{split}
\eeq
where initial states $\o_0$ and $\tilde{\o}_0$ are
independently, randomly chosen and
$\|\theta_1-\theta_2\|$ denotes the
 (shortest) distance between two
  states $\theta_1$ and $\theta_2$. Figure~\ref{fig:dist} (A) illustrates
these measurements.  Both $X(t)$ and $Y(t)$ denote the
distance between a pair of trajectories at time $t$:
``within-ensemble" for $X(t)$ and ``between-ensemble" for $Y(t)$. Formally, they are
random variables because they depend on a pair of
random initial conditions.
   %@@@@@@@@@@@@@@@@@@@@@@@@@@@
%@@@@@@@@@@@@@@@@@@@@@@@@@@@

The quantity $X(t)$ represents the ``typical
size'' of an ensemble at time $t$; we view it like the
diameter in that it measures the distance between two
typical points on the corresponding attractor.
The quantity $Y(t)$ is the typical distance
between two ensembles elicited by stimuli $I^A(t)$ and
$I^B(t)$. In the context of the present model, we say
that the two stimuli are {\it separable} (in the state
space sense) if the distributions of $X(t)$ and $Y(t)$
do not have a significant overlap (see Figure~\ref{fig:dist}).
The dimension of the underlying attractor is
  reflected in $X(t)$ in the following way: if the
  dimension were 0 (so the attractor is a single
  point), then $X(t)\approx0$, and if the dimension is
  positive, then $X(t)$ would also be positive (but
  can be large or small). In general,
  $X(t)$ will fluctuate in a complicated fashion as
  the time-dependent attractor evolves and changes
  shape.  Based on previous work~\cite{Lajoie:2013p17297,Lajoie:2014p18333}, we expect
  these fluctuations to be relatively small, so that
  $X(t)$ is nearly constant in time.
 Furthermore, while the fluctuations in $X(t)$
  depend on the specific choice of stimulus, standard
  results from probability theory tell us that its
  mean depends only on the system parameters and the
  statistical distribution of the stimuli.  Thus, the
  mean value of $X(t)$ remains unchanged if we
replace $I^A(t)$ by $I^B(t)$ in
Equation~\eqref{dist_RV} (since we require $I^A(t)$
and $I^B(t)$ to have identical parameters). Similar
  reasoning applies to $Y(t)$, which is akin to
  measuring the rate of separation between two
  independent realizations of the same stochastic
  process. Once again, we expect $Y(t)$ to
  fluctuate around some mean, with the exact
  fluctuations dependent on the input realizations
  but with mean and variance depending only on system
  parameters and stimulus statistics.

From simulations, we produce pairs of trajectories
subject to both $I^A(t)$ and $I^B(t)$, all with random
initial states. We compute $X(t)$ and $Y(t)$ and plot
the result as a function of time in
Figure~\ref{fig:dist} (B) where $\r_{corr}=0.5$ ($I^A$
and $I^B$ are $50\%$ correlated). As expected, after a
short transient $X(t)$ settles quickly to a positive
  constant, which is unchanged if the trajectory pair
is selected from $\O_{I^A}^t$ or from $\O_{I^B}^t$. Likewise, $Y(t)$ settles quickly to a steady-state in which it fluctuates around a well-defined mean.

For a more complete view of this phenomenon, we
consider the distributions of pairwise distances,
sampled over 100 trajectory pairs, starting from
uniformly random states. Figure~\ref{fig:dist} (C)
shows the time evolution of these distributions for
the first second of elapsed time.  As for the single-pair measurements
(Figure~\ref{fig:dist} (B)), both distributions settle
into near-constant,
  steady values after a very fast transient
($\sim$10-50 ms). This remains true for any similarity
parameter value $\r_{same}$, $\r_{corr}$. We note that
this short transient validates our general definition
of ``trial" which includes trajectories with distinct
initial states as well as any trajectories that
received some sort of perturbation --- such as a
synaptic failure or the event of an extra spike --- in
the recent past. To capture the stationary nature of pairwise distances, we compute time-averaged
distributions $\langle X\rangle_t$ and $\langle
Y\rangle_t$, which we calculate using $1000$ time points
between $t=100$ and $t=12,000$ ms.  While $\langle
X\rangle_t$ remains the same regardless of the similarity
between $I^A(t)$ and $I^B(t)$, $\langle Y\rangle_t$
can be used to measure the effect of stimulus similarity
on the location of response ensembles in state space.
Figure~\ref{fig:dist} (D) shows the means $\langle
X\rangle_t$ and $\langle Y\rangle_t$ as well as one
averaged standard deviation, for a range of input
similarity parameters $\r_{corr}$ and $\r_{same}$
between $0$ and $1$. As expected, when both inputs are
identical ($\r_{corr}=\r_{same}=1$), $\langle
Y\rangle_t$ collapses to $\langle X\rangle_t$ since
$\O_{I^A}$ and $\O_{I^B}$ describe the same
ensemble. 

However, we see that $\langle X\rangle_t$
and $\langle Y\rangle_t$ become more than two standard
deviations apart as soon as the stimuli become less than $90\%$ similar.  This is true for both definitions of stimulus similarity.  The conclusion is that the chaotic, balanced networks at hand produce dynamical responses that stay separated in state space even for stimuli that are very similar.

\smallskip
To summarize the above, pairwise distances between evoked
trajectories inform us about geometric properties of
response ensembles, and how they organize in state
space. In parameter regimes where $X\ll Y$, which occur for stimuli that are  less than $90\%$ similar, we expect that a decoder could classify novel
trajectories, given information about the
distributions $\langle X\rangle_t$ and $\langle
Y\rangle_t$.  Conversely, in parameter regimes
  where $X\approx Y$, it may be difficult for a
  downstream decoder to discriminate response
  ensembles using simple criteria like linear
  separability.  We now investigate this stimulus discrimination in detail.

%---------
\subsection{Finding 1.~Chaotic spike patterns are
  linearly discriminable}

The state-space separability studied above assumes that one
  has access to the full state of the network
  at all times; any biologically realistic
  decoder would only have access to a network's spiking activity.  We next present evidence that the {\it spikes} generated by balanced, chaotic networks can also be used by a simple linear classifier, the
  Tempotron, to discriminate between two stimuli with
  identical statistical properties.  We first review 
  some general findings about spike responses in chaotic networks.

%-----
\subsubsection{Trial-to-trial variability of single
  cells and reliable spike events}

For the class of models studied here, previous work
has shown that the unstable directions in attractors -- i.e., the directions in which chaotic dynamics will spread the response ensemble produced by a single stimulus -- 
generally align with neural coordinates.  Moreover, the
identity of the corresonding ``unreliable" neurons change in time~\cite{
  Lajoie:2013p17297}. This leads to intermittent
variability in single neurons: at any moment there is
a small fraction of neurons
in the network that have variable dynamics across
trials, while the rest behave in a reliable fashion.

We can observe this phenomenon by restricting the
definition of pairwise distance~\eqref{dist_RV} to
single-neuron coordinates: $X_i(t)$, and $Y_i(t)$ for the (transformed) voltage variable $\o_i$. The value $X_i(t)=0$   means that different initial network states 
  nevertheless lead to the same state for cell $i$ at
  time $t$, i.e., the neuron $i$ has the same voltage
  across all trials.  A similar interpretation applies to
  $Y_i(t)$. Figure~\ref{fig:dist} (E) shows these
quantities for a
randomly selected neuron $i$. In contrast to pairwise
distances in the full state space (Figure~\ref{fig:dist}
(B)), $X_i(t)$ regularly collapses to zero.  This is a reflection of the
intermittent variability discussed above. $Y_i(t)$ also varies
over time although it remains greater on average. This
suggests that at any
moment in time, for a given input $I(t)$, some network
coordinates may offer trial-to-trial reliable
responses that can be used to distinguish similar input stimuli.

%@@@@@@@@@@@@@@@@@@@@@@@@@@@
 \begin{figure*}[h!]
%\begin{center}
\includegraphics{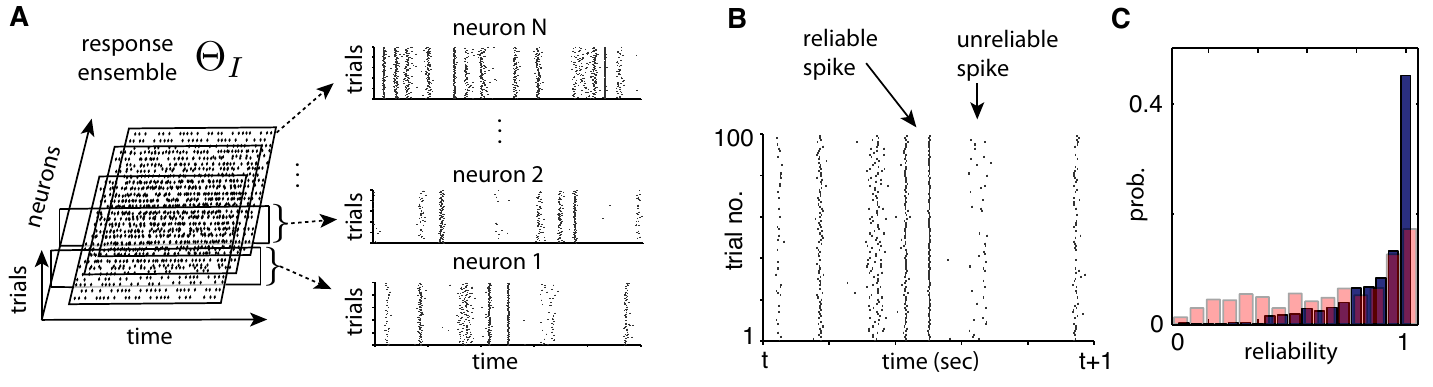}
\caption{({\bf A}) The spike patterns from a response ensemble, $\O_I$, containing network responses to input $I(t)$ on many trials, are separated for different neurons. This leads to a collection of $N$ single-neuron rasters across trials.
  ({\bf B}) Example raster plot for a single neuron in the network on
  one hundred trials. Some spikes are precisely
  repeated across all trials (reliable) while some
  others are variable from trial to trial
  (unreliable).
  ({\bf C}) Histogram of spike
  event reliability sampled over all spike events in
  the network (dark blue), and over uncoupled neurons
  where network interactions are replaced by
  trial-independent poisson spike trains (light
  red). See text for details, one hundred trials used
  for sampling.
 }
\label{fig:spike_rel}
%\end{center}
\end{figure*}
%@@@@@@@@@@@@@@@@@@@@@@@@@@@

We now turn our attention from the internal
  states of cells to the spikes produced by response
ensembles $\O_I^t$. This is a collection of
events indexed by time, neuron and trial as
illustrated in Figure~\ref{fig:spike_rel} (A) and
denoted by $\O_I$ (without explicit time-dependence $t$).
Consistent with the intermittent trial-to-trial
variability in the state variables of single cells, we find that the spiking of single cells shows changing levels
of reliability (i.e., repeatability across trials).
Figure~\ref{fig:spike_rel} (B) shows the spikes
generated by a randomly chosen neuron in the network over 100 repeated trials.  It is clear that
some spikes are repeated across all trials ({\it
  reliable} spikes) and others are not ({\it
  unreliable} spikes).  We quantify the {
  reliability} of a spike by estimating the
probability of it being repeated on other trials. This
is done by convolving the cross-trial spike trains of
a neuron with a gaussian filter (standard
deviation 10 ms) and adding the resulting waveforms
to obtain a filtered peri-stimulus time histogram
(PSTH). We call each peak in this histogram a {\it
  spike event} and a spike is assigned to an event if
it falls within a tolerance time-window defined by the
width of the peak at half
height~\cite{Fellous:2003p16226,Lajoie:2013p17297}. The
event's {\it reliability} is then estimated by the
number of member spikes divided by the number of
trials considered. If a spike event has a reliability
of 1, it means that one can expect a spike from that neuron at that
moment on every trial while a lower reliability
indicates more variability.

Importantly, despite chaotic dynamics, a majority of
spike events in the network are reliable, as
illustrated by the histogram (dark blue) in
Figure~\ref{fig:spike_rel} (C). This is due to the
geometric properties of response ensembles described
above and further outlined
in~\cite{Lajoie:2013p17297,Lajoie:2014p18333}.  A
consequence is that at any one moment, there typically is reliable spiking in some neurons'
coordinates (see Figure~\ref{fig:spike_rel} (A)),
leading to repeatable spike patterns embedded within
population patterns.
 
As demonstrated in  detail in earlier work, we stress that reliable spike events are not solely the result of strong fluctuations of inputs $I_i(t)$ driving neuron $i$ to spike reliably. While input fluctuations certainly contribute to spiking events, so does the timing of incoming synaptic events.  To illustrate this, Figure~\ref{fig:intro} (D) shows a ``surrogate" histogram of spike event reliability (light red), in which neurons are driven by the same input signals but network interactions are replaced by trial-independent, poisson-generated spike trains matching the network firing rate. It is apparent that replacing chaotic variability by homogeneous poisson randomness makes spikes become much less reliable (see also~\cite{Lajoie:2013p17297}). We revisit the roles of external inputs and network interactions in spike generation in the section {\it Signal strength modulates a noise floor from chaos} below (see e.g. Figure~\ref{fig:parametric} (A)).

\smallskip
These results
  indicate that the response ensembles $\O_I^t$ elicited by any
fixed input $I(t)$ correspond to relatively
low-dimensional attractors whose projections onto the
neural coordinate axes (i.e., the single-cell marginal
probability distributions) are tightly distributed, except
for a few directions at a time. These directions will
change with time as the input $I(t)$ fluctuates, but
the number of unstable directions remains roughly
constant. This leads to changing trial-to-trial
variability of spike times in single neurons. Intriguingly, similar types of
``intermittent" spiking variability have been reported
in {\it in vivo} experiments
(cf.~\cite{Tiesinga:2008p12846,Reinagel:2000p11391}).

%---------
\subsubsection{The Tempotron and discrimination}

Our goal is to discriminate between two
  statistically identical random stimuli based on the network
  responses they evoke.  We do this by training a
classifier on the collections of spike times $t=t^i_s$
evoked by distinct network inputs $I^A(t)$ and
$I^{B}(t)$ on finite time intervals $[0,T]$.  There
are many machine-learning techniques that can perform
this task, but our main criteria for a preferred
approach are: (i) it should be a useful metric to
compare the encoding performance of our network under
different conditions and (ii) it should isolate
important spike features for coding
(interpretability).  We therefore opt for a 
simple, linear classification approach. The results serve as a lower
bound on the classification capacity of the network.

There several approaches one can use to find a hyperplane that separates sets
of points in a high-dimensional space, such as the
Support Vector Machine~\cite{Ambard:2012p952} or other
regression techniques. Here, we use a classification
method called the {\it
  Tempotron}~\cite{Gutig:2006p10816}, a
gradient-descent approach acting on linear weights of
temporal kernels designed to mimic the post-synaptic
potentials induced by individual spikes. Importantly, the resulting classification hyperplane corresponds to the
threshold of a spiking Linear Integrate-and-Fire (IF)
neuron model.

The Tempotron receives vector-valued filtered spike
trains $s(t)=\{s_i(t)\}_{i=1...N}$ where
$s_i(t)=\sum_{t^i_s} K(t-t^i_s)$ and
$K(t)=V_0[e^{-t/\tau_1}-e^{-t/\tau_2}]$ where $V_0$ is
a normalizing constant.  The double-exponential
filtering is meant to mimic the rise and fall of
synaptic potentials in an IF neuron whose
voltage obeys the equation
\beq
V(t)=\sum_{i=1}^Nw_i\sum_{t^i_s}K(t-t_s^i)+V_{rest}
\label{eq:tempo}
\eeq
with voltage threshold set at $V_{thr}=1$.
Thus, the Tempotron computes the sum
of the filtered network spike trains, according to the
read-out weights $w_i$ and the timescale set by its kernel.
We tune the filter's
decay and rise time-constants to $\tau_1=20$ and
$\tau_2=3.75$ ms as in~\cite{Gutig:2006p10816}, to impose an intrinsic sensitivity to spike timing at that resolution.

The Tempotron's goal is to fire at
least one spike (i.e. cross its threshold $V_{thr}$)
when presented with a network spike output associated
to $I^A(t)$ while refraining from firing when the network
responds to $I^B(t)$. 
Following~\cite{Gutig:2006p10816},
we train the Tempotron to classify spike outputs on
finite-length time intervals $[t_0,t_0+T]$ using a
fixed number of trials from $\O^t_{I^A}$ and
$\O^t_{I^B}$ respectively, and test the trained
classifier on new trials. Thus, beyond 
discriminating between two ``training" ensembles of spikes, we test the ability of the Tempotron to
generalize and discriminate new patterns, related to training sets in that they are sampled from the same chaotic response ensemble. The robustness of the Tempotron to Gaussian, random spike-time jitters is well established~\cite{Gutig:2006p10816}; here we investigate the effect of chaotic variability.   
%@@@@@@@@@@@@@@@@@@@@@@@@@@@
 \begin{figure*}%[h!]
%\begin{center}
\includegraphics{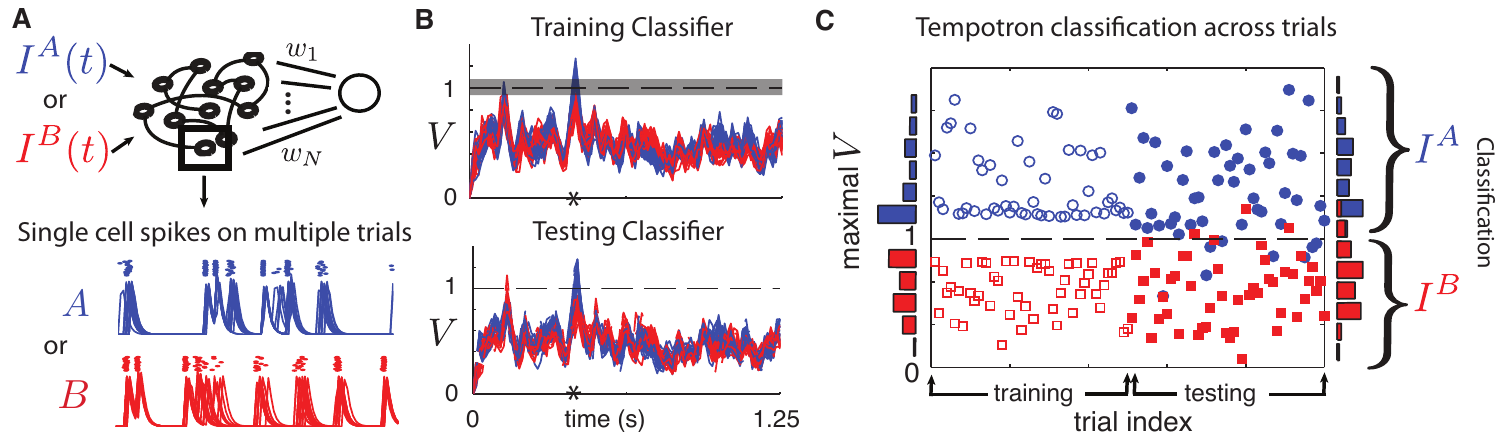}
\caption{({\bf A}) Schematic of a Tempotron with readout weights $w_i$.  For one neuron from the chaotic network, we show spike times (dots) of a randomly chosen neuron across several trials for inputs $I^A$ or  $I^B$, together with the filtered traces for these spike
  times. ({\bf B}) Tempotron voltage traces $V(t)$ for a time window of $T=1.25$
  seconds. Top: traces used for training. Bottom:
  traces used for testing. Grey shaded area
  surrounding threshold $V_{thr}$ shows training
  margin. Asterisk shows time of maximal output. ({\bf C}) Classification for all
  training and testing trials. Markers show the
  Tempotron's maximal voltage $V$ in time
  window $T$. A correct classification corresponds to blue
  circles ($I^A$) above threshold and red squares
  ($I^B$) below.  For all panels, the stimuli $I^A$ or $I^B$ have $\r_{same}=0.9$.}
\label{fig:tempo}
%\end{center}
\end{figure*}
%@@@@@@@@@@@@@@@@@@@@@@@@@@@

Figure~\ref{fig:tempo} illustrates the process.  Out of $100$ trials from each
ensemble, we select $50$ from $\O_{I^A}$ and $50$ from
$\O_{I^B}$ to train the read-out weights
$w_i$ in
Equation~\eqref{eq:tempo}. Figure~\ref{fig:tempo} (A)
illustrates the filtered spike output of a randomly selected
neuron on the training trials from $I^A$ and $I^B$. The remaining trials will be used for
testing. We employ the algorithm described
in~\cite{Gutig:2006p10816} to find a set of weights
$w_i$ imposing that the voltage $V$ of the Tempotron
will exceed the threshold $V_{thr}=1$ at least once
in the pattern time-window when presented with spikes from a trial
from $\O^t_{I^A}$ while remaining sub-threshold when
receiving spikes from $\O_{I^B}$. We report results for a time window of $T=1.25$ seconds. We find that training often fails for $T<50$ ms, but that the results we present below remain qualitatively unchanged for bigger $T$.
 We use a
margin of $V_{thr} \pm 0.1$ in the
training to ensure a good
separation (c.f.~~\cite{Gutig:2006p10816}). Figure~\ref{fig:tempo} (B) shows the
output of the Tempotron during training and testing
while panel (C) compares the training and testing
outcomes by showing the maximal $V$ within the
$T$-window. In this example, even though inputs are quite similar 
($\r_{same}=0.9$), only a few test trials are misclassified.

To quantify the discriminability of spike patterns, we
define the {\it performance} $P$ as
the fraction of successful test classifications. Note
that $P$ has a maximal value of $1$ and a minimal
value of $0.5$ which corresponds to chance. We average
$P$ by retraining the Tempotron $20$ times using
different training and testing trials from our
ensembles. As an example, the performance $P$ of the
classification in Figure~\ref{fig:tempo} (C) is about
$0.9$. 

Our overall finding is is that $P$ roughly follows the trends
found above for the pairwise distance
between response ensembles (see Figure~\ref{fig:dist}): we observe perfect
classification ($P=1$) until input similarity reaches about $90\%$ (i.e. $\r_{same}$, $\r_{corr}=0.9$).  
This means there exists a linear combination of evoked spike patterns that reliably sum to cross a threshold for only one of the two stimuli, regardless of network initial conditions.
	{\it Thus, when distinct stimuli are presented to chaotic networks, even ones with very similar features, it is not only the network states they produce that are highly discriminable, but also the resulting spike trains.}
We will return to the question of performance versus 
input similarity below, but first turn to its mechanism.  

%---------
\subsubsection{Reliable spiking leads to discriminability}

We propose that reliable spiking of a few neurons at
precise moments in time (i.e., reliable spike patterns) drive successful classification of stimuli by the Tempotron.  In this section we demonstrate this by deconstructing the Tempotron's readout and by observing the impact of ``jittering" underlying spikes.

%@@@@@@@@@@@@@@@@@@@@@@@@@@@
 \begin{figure*}[h!]
%\begin{center}
\includegraphics{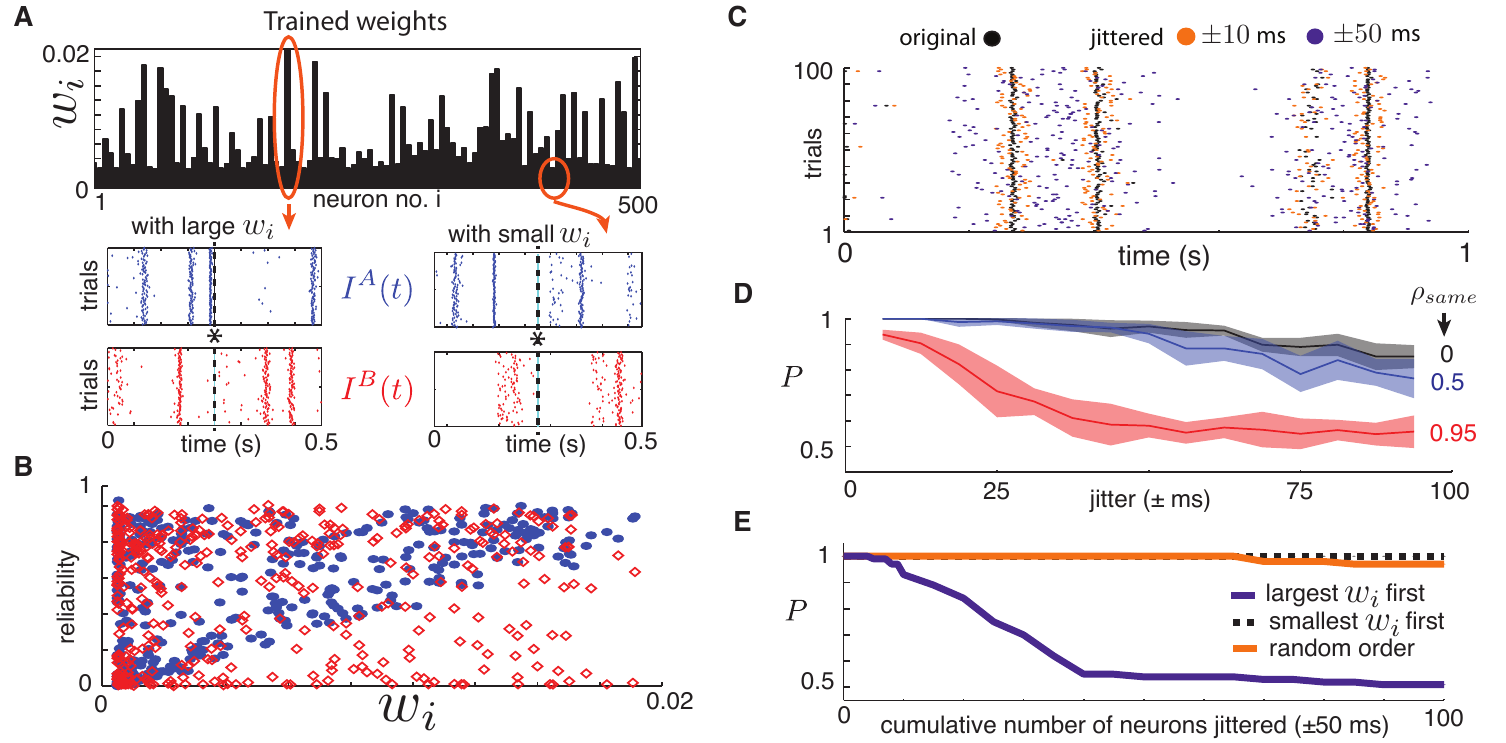}
\caption{({\bf A}) Top: Trained Tempotron weights
  $w_i$. Bottom: example cross-trial spiking output
  for neurons associated with large and small weights
  $w_i$.  Asterisk shows the time at which the maximal value of the Tempotron variable $V(t)$ is obtained
  (see also Figure~\ref{fig:tempo} (B)).
   ({\bf B})
  Spike-time reliability for the closest spike to the maximum (asterisk), plotted for each cell 
  against the read-out weight $w_i$. Solid blue markers
  indicate responses to $I^A(t)$ and hollow red
  markers to $I^B(t)$. ({\bf C}) Example raster plot
  for a single neuron across 100 trials, with and without
  spike-time jittering.  Two jitter strengths shown: $\pm 25$ and $\pm 100$ ms. ({\bf D}) The Tempotron's performance $P$ when all neurons are jittered, plotted against jitter strength, for three values of $\r_{same}$. The classifier was trained and tested on jittered spike time data.
  ({\bf E}) Tempotron's performance $P$ against the
  cumulative inclusion of neurons where the spike
  times were strongly jittered ($\pm 100$ milliseconds). Different lines indicate three
  orderings of neurons the neurons that were jittered, based on their
  trained read-out weight $w_i$: (i) decreasing, (ii)
  increasing and (iii) random. The classifier was trained on original data and tested on jittered data from a network of $N=500$ neurons.  For all panels except (D), $\r_{same}=0$.}
\label{fig:jitt}
%\end{center}
\end{figure*}
%@@@@@@@@@@@@@@@@@@@@@@@@@@@

We first turn to the read-out weights
$w_i$, which are the result of a global optimizing algorithm~\cite{Gutig:2006p10816}. Figure~\ref{fig:jitt} (A) shows the values of all
$w_i$'s for a Tempotron trained to distinguish two stimuli, $I^A(t)$ and $I^B(t)$, with perfect
performance ($P=1$).  We find that the neurons with the strongest weights spike very reliably when the network is
presented with input $I^A(t)$, right before the peak
of the Tempotron's output (asterisk, see
Figure~\ref{fig:jitt} (A) bottom and Figure~\ref{fig:tempo} (B)). Conversely, the
same neurons either do not spike or spike unreliably in
response to $I^B(t)$ around the same moment in time. In
Figure~\ref{fig:jitt} (B), we quantify this finding by
showing a scatter plot of spiking event reliability
for each neuron, plotted against
its output weight $w_i$.  This shows that any neuron
with a high $w_i$ also shows high spiking reliability
under input $I^A(t)$.

 Next, we degrade the temporal precision of network responses by randomly ``jittering" spike times and study the impact on classification performance by the Tempotron. To ``jitter" a spike train, we shift each spike time on all trials by a random amount, uniformly and independently
drawn from an interval of $(-r,r)$, where $r$ is the {\it jitter radius} (see Figure~\ref{fig:jitt} (C)). This leaves the total number of spikes fired the same, but strongly disrupts their temporal precision, as illustrated in Figure~\ref{fig:jitt} (C).
We use this jittering in two ways.  First, we train and test the Tempotron on jittered spikes, to probe the dependence of classification performance on the temporal precision of spike times produced by chaotic networks.  Second, we train the Tempotron on the original spike data but test using jittered spike times for subsets of neurons, to probe the learned role of these neurons in classifying stimuli.  
  
Training and testing the Tempotron on jittered spike time data shows that classification performance $P$ progressively declines as the jitter radius increases. The rate at which performance drops depends on the similarity between inputs $I^A(t)$ and $I^B(t)$, as illustrated in Figure~\ref{fig:jitt} (D) for three values of $\r_{same}$ (the fraction of neurons receiving identical direct inputs under both signals).  Evidently, the lower $\r_{same}$ is, the more distinguishing features there will be in the two response ensembles to be classified, the combination of which enables the Tempotron to classify stimuli even  with substantial spike time jitter.  Overall, when spikes are jittered by 10's of ms, classification performance drops significantly; for similar stimuli, performance drops halfway to chance when the jitter radius is 25 ms.  We conclude that the Tempotron uses quite precisely timed spikes to distinguish the responses of chaotic networks to nearby stimuli.
 
To probe the question of what spiking features lead to the reliable classification learned by the Tempotron, we set the jitter radius to $50$ ms and jitter different subsets of neurons in the testing data.   
In contrast to the procedure described above, where training and testing spike times were jittered, only testing spike times are now jittered; the Tempotron is trained on the original spikes produced by the network.
We apply this jittering
  procedure to an increasing number of neurons in the
  network, re-testing the classification performance
  $P$ as neurons are cumulatively added to this
  ``jittered" pool.  We do this for three different
  orderings in which neurons are added to the jittered pool: (i) neurons with largest
  trained weights $w_i$ listed first, (ii) neurons
  with smallest $w_i$'s listed first and (iii) neurons
  randomly listed.  The results are shown in
  Figure~\ref{fig:jitt} (E).  First, note that perturbing the spike times of neurons with large
  $w_i$'s quickly reduces performance down to chance,
  whereas jittering the spike times of random neurons
  or those with low read-out weights has little to no
  effect on performance. The details of the observations above are likely to depend on the choice of time constants for the Tempotron's filters, but we expect the overall trends to persist for a range of these constants, based on the spike-time reliability described earlier.  From this we conclude that the Tempotron learns to classify stimuli based on the reliable spikes of a few neurons, at precise moments in time.
  
Finally, we shuffled the spike ensembles across trials by building
surrogate spike patterns in which the spike output of each neuron is
taken from a randomly chosen trial from the ensemble; if the read-out
cells that serve as strong inputs to the Tempotron (i.e., the cells with
relatively large $w_i$) were unreliable, this would have the effect of
shifting their spike times and changing the spike counts within the test
window.  Numerical results indicate that shuffling has no
appreciable effect on classification performance. This further suggests that
repeatable spike patterns are responsible for good classification,
rather than statistics like spike counts on longer time scales.
\smallskip

Taken together, these tests show that {\it despite
  chaos, the network response preserves a fair degree
  of spike reliability across trials in key neural
  coordinates, and a simple decoding scheme is capable
  of taking advantage of this reliability to accurately classify spike trains.}  Note that the
identity of these ``key" neurons will change for different input stimuli 
and thus, at least in principle, many readout schemes could be trained
in parallel on the same network.

%------------------------
\subsection{Finding 2.~Recurrence enables information distribution within networks, discrimination using few read-outs}

We have shown that a
neural-like readout, the Tempotron, can use the reliable
spikes embedded in a chaotic network's response in order to classify input stimuli. Up to now, this classifier had access to spikes from every neuron.  A natural question is whether this complete access is necessary.  In other words, how does classification performance depend on the number of inputs and outputs
to the network? For example, if only a few neurons
in the network receive discriminable inputs
($I^A_i(t)  \neq I^B_i(t)$) and the decoder only reads
out from a few (different) neurons, do network
interactions distribute enough stimulus information to enable a successful
classification?

%@@@@@@@@@@@@@@@@@@@@@@@@@@@
 \begin{figure*}[h!]
%\begin{center}
\includegraphics{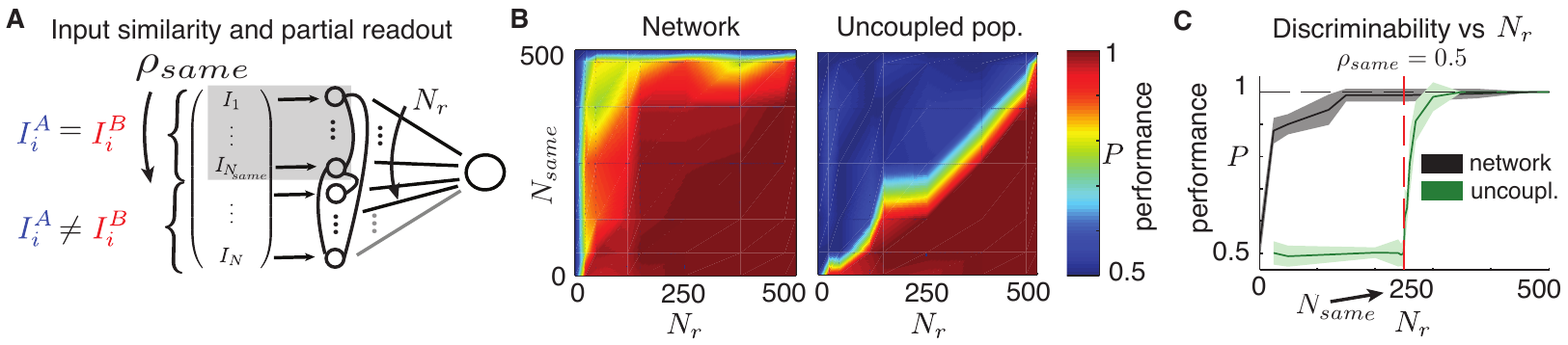}
\caption{({\bf A}) Illustration of input similarity $\r_{same}$
  and partial read-out dimension $N_r$. The first $N_{same}$ neurons receive the same direct inputs
  from $I^A$ and $I^B$ and the Tempotron only reads
  spikes from the first $N_r$ neurons. ({\bf B})
  The Tempotron's discrimination
  performance $P$ as a function of $N_r$ and
  $\r_{same}$ for the chaotic network (left) and populations of
  $N$ uncoupled noisy neurons (right). Performance
  obtained using ensembles of $100$ trials from each
  input evenly separated into training and testing
  sets. Values of $P$ reported are the averages of
  $20$ training cycles using different training and
  testing subsets from the ensembles. ({\bf C})
  Performance vs $N_r$ profile for both the network and the
  uncoupled population with $\r_{same}=0.5$ ($N_{same}=250$). Shaded
  areas show the standard deviation of $P$.}
\label{fig:class}
%\end{center}
\end{figure*}
%@@@@@@@@@@@@@@@@@@@@@@@@@@@

In our model,
the proportion of the input that is directly discriminable is controlled by the
parameter $\r_{same}$ -- the fraction of cells that
receive identical inputs under $I^A(t)$ and $I^B(t)$. We adopt an enumeration that lists these cells first for ease of notation: the first $N_{same}$ neurons receive indistinguishable inputs while the remaining $N_{diff}$ receive independent ones for each stimulus; with $N_{same}=\r_{same}N$ and $N_{diff}=N-N_{same}$ (see Figure~\ref{fig:stims} (B)). In addition, we introduce a second parameter, $N_r$, which controls the number
of read-out neurons providing inputs to the Tempotron. We impose the same ordering so that the first $N_r$ neurons will be read out (i.e. the Tempotron weights $w_i$ are defined for indices $i=1,...,N_r$) as depicted in
Figure~\ref{fig:class} (A). 
In this way, the classifier always reads out neurons that receive identical inputs, and when $N_r<N_{same}$, it only has access to such neurons.

Classification performance $P$ as a function of $N_{same}$ and $N_r$ is plotted in
  Figure~\ref{fig:class} (B). Consistent
 with the results above, 
 we achieve
  near-perfect performance ($P \approx 1$) up to
  $\r_{same}\sim0.9$ when reading out from the full
  network $N_r=N$. However, we also obtain
  very high classification performance in many cases where
  $N_r<N_{same}$.  Here the classifier can {\it only}
  read out from neurons that receive identical inputs under the two stimuli (i.e. not directly discriminable).  This means that discriminable features of these neurons' spike patterns cannot originate from their direct inputs $I_i^A(t)$ and $I_i^B(t)$, and must instead result from network interactions that communicate activity from other neurons.
  
To help put these results in context, consider what would happen for a population of $N$ uncoupled cells receiving the same inputs as the network.  In the uncoupled population of QIF neurons, there cannot be
  any chaotic behavior:  a general fact is that isolated QIF neurons are
  always reliable~\cite{Lin:2009p822}, meaning that their
  response ensembles collapse to single trajectories
  after a short transient.  This means that by reading
  out from the entire population, one could classify inputs that share all but one
  direct input $I_i(t)$ (i.e. $N_{same}=N-1$).
  However, if a classifier only had access to a few
  neurons that do not receive discriminable 
  stimuli, classification would fail completely.
This is demonstrated in Figure~\ref{fig:class} (B).  In these numerical
  simulations, we added independent gaussian noise to each neuron as a surrogate for chaotic variability:  we choose the amplitude to match the $X$ distance statistics from
  Equation~\ref{dist_RV} to values for chaotic networks given in
  Figure~\ref{fig:dist} (B). As
  expected, when $N_r<N_{same}$, the uncoupled
  population has a classification performance of $P=0.5$, which is
  random chance.  When $N_r > N_{same}$, $P$ rapidly transitions to nearly perfect performance.
    
 Figure~\ref{fig:class} (C) displays a cut through the previous panels at $\r_{same}=0.5$, showing
 that the recurrent, chaotic network outperforms the
uncoupled population for all $N_r$.  Thus, classification is more robust to the number and identity of cells that are read out for recurrent, chaotic networks that for noisy, uncoupled ones. We also repeated this procedure using randomly selected read-out neurons $N_r$ (violating the ordering described above) which included both neurons that received
distinct stimulus inputs and those that did not. In these
cases, the recurrent networks continued to outperform noisy uncoupled populations, but by a smaller margin
(data not shown). We also tested classification when adding the same amount of gaussian
noise to the recurrent networks as for the uncoupled populations, and found that recurrent networks still showed better performance when $N_r<N_{same}$ and remained
comparable to the uncoupled population otherwise (data not
shown).  This means that even with a double source of
variability (noise and chaos), recurrent networks still permit a high level of stimulus classification.

\smallskip

We conclude from these results
  that {\it recurrent interactions of the type found in our network model, despite generating chaotic
  instabilities, are an effective and robust way to distribute information about local inputs throughout a network so that distant  readouts can be used for stimulus discrimination without specifying precise 
  connectivity.}  We expect these findings to hold generally in
  recurrent networks that are well-connected, in the sense that for every pair
  of vertices in the associated connectivity graph,
  there is a relatively short directed path between
  them.  In the next section, we show how the
relationship between the number of readouts $N_r$
needed for classification and the number of distinct
inputs $N_{diff}$ depends on stimulus statistics.

\bigskip

%---------
\subsection{Finding 3.~Signal strength modulates a ``noise floor" from chaos}
\label{section:params}

So far, our investigation of stimulus encoding has been restricted to a
single choice of the stimulus amplitude
$\e$ and mean $\eta$ ($\e=0.5$, $\n=-0.5$).  For these parameters, we showed that even highly similar stimuli can be distinguished based on the responses of chaotic networks.   How does this depend on the stimulus amplitude (i.e.~strength of temporal features)?  When this amplitude drops, one might expect that it will eventually fall below a limit when any differences in stimuli will be obscured by variability induced by the chaotic dynamics of a network.  We call this limit the chaos-induced {\it noise floor}.  Below, we study this noise floor, and thereby establish how the discriminability of stimuli in chaotic networks depends on their statistics.\bigskip
  
\begin{figure}[h!]
%\begin{center}
\includegraphics{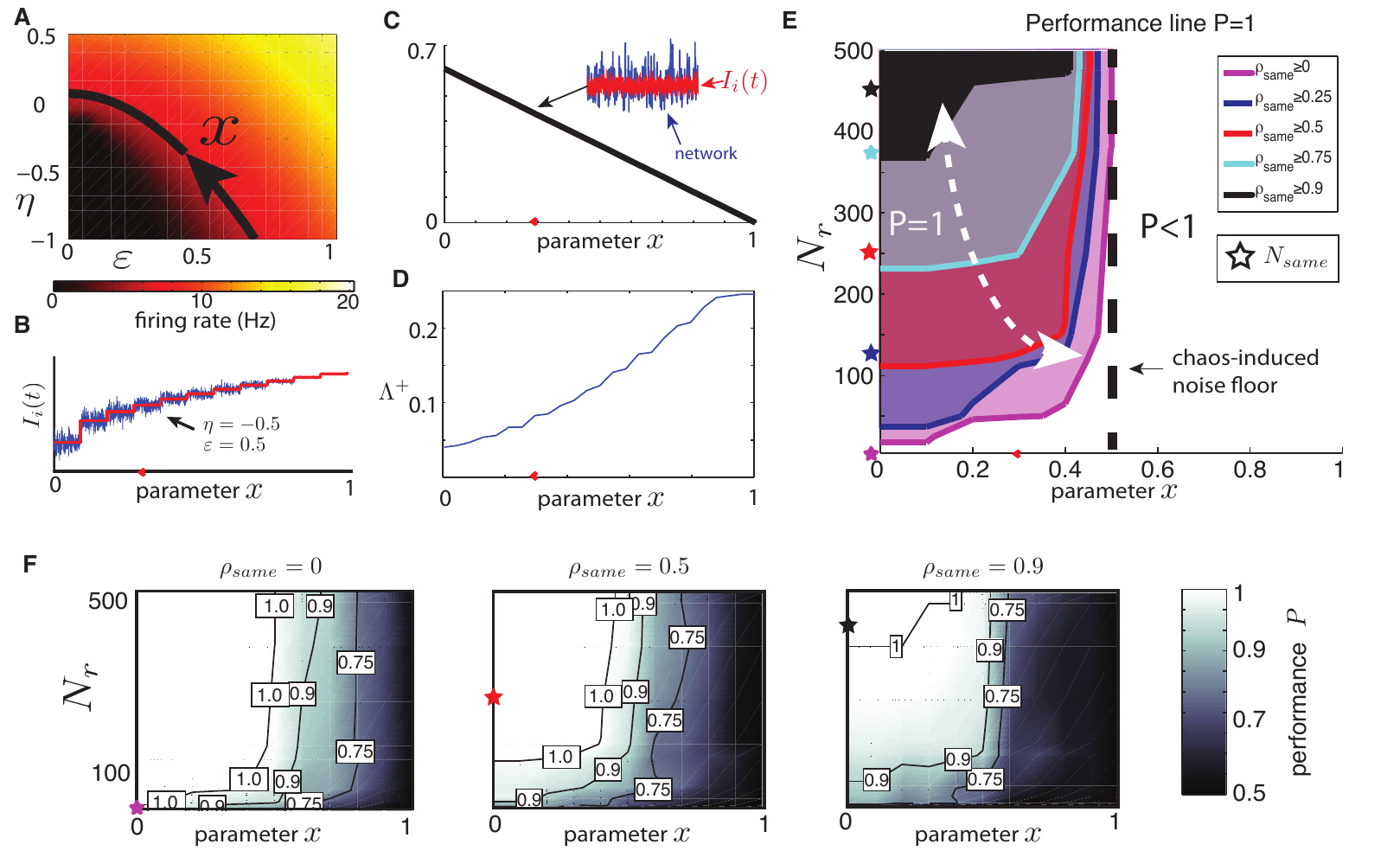}
\caption{ ({\bf A}) Mean firing rate of
  the network as a function of mean input $\n$ and
  signal amplitude $\e$. Black curve shows level set
  at 6.5 Hz, and is parametrized by the normalized
  arclength $x$ in direction of arrow. Parameter set
  used throughout the rest of paper ($\e=0.5$,
  $\n=-0.5$) corresponds to $x\simeq 0.3$.  ({\bf B})
  Illustration of input $I_i(t)$ presented to a neuron
  as a function of the parameter $x$. The red line
  shows the mean $\n(x)$ while the blue line shows
  fluctuations of amplitude $\e(x)$. For all
  parameters, the mean excitatory firing rate is
  constant.  ({\bf C}) Ratio of the amplitude $\e$ of $I_i(t)$ to that of the standard deviation of network interactions as a
  function of parameter $x$ (computed in $\o$
  coordinates and integrated over increments of 1.25
  milliseconds). Inset shows input (red) and network
  interactions to a single neuron for our benchmark parameters
($\e=0.5$, $\n=-0.5$).  ({\bf D}) The
  fraction of positive Lyapunov exponents
  $\#\{\l_i>0\}/N$ as a function of the parameter
  $x$. Red dot shows the benchmark regime
  ($\n=-0.5$, $\e=0.5$).  ({\bf E})
  Level curves of classification performance $P=1$ in
  the $(x,N_r)$ parameter space for several values of
  $\r_{same}$. For parameter pairs to the left of these
  curves, the network achieves perfect
  classification. Stars indicate the corresponding
  $N_{same}=\r_{same}*N$.  ({\bf F}) Classification performance $P$ as a function of $x$
  and read-out dimension $N_r$ for three input similarity
  fractions $\rho_{same}=0,\,0.5, \,0.9$. Stars
  indicate the corresponding $N_{same}=\r_{same}*N$.}
\label{fig:parametric}
%\end{center}
\end{figure}

To systematically compare how stimulus statistics impact discriminability, care is needed to keep the network activity in a fairly consistent firing regime.  We do this by varying parameters $(\e, \; \n)$ together in a way that will produce a fixed firing rate, averaged across the network. This way, we can be certain that classifiers will be trained on the same number of spikes on average, a quantity that could affect the interpretation performance $P$ if left uncontrolled. Specifically, we vary $\eta$ and
  $\e$ together along a path that leaves the
  network-wide average firing rate fixed at 6.5 Hz, as was the case for parameter values used above (see
  also~\cite{Lajoie:2014p18333}).  As illustrated
    in Figure~\ref{fig:parametric} (A), we
  parameterize this path by a normalized arclength
    parameter $x$: for higher values of $x$,
  $\n$ is larger and $\e$ is smaller (see Figure~\ref{fig:parametric} (B) for illustration of input as $x$ changes). 
  
   %@@@@@@@@@@@@@@@@@@@@@@@@@@@
%%@@@@@@@@@@@@@@@@@@@@@@@@@@@

We begin by visualizing the relative amplitudes of external inputs $I_i(t)$ and network
interactions. If the strength of stimulus fluctuations $\e$ was so
great as to simply overwhelm interactions between neurons
in the network, then the role of chaos in any conclusion about stimulus encoding would be trivial:  the stimulus simply overwrites any intrinsic dynamics.  Figure~\ref{fig:parametric} (C) shows this is not the case, neither for our benchmark parameters nor for the others we study below.  
For a range of $x$ values, we plot the ratio of fluctuation amplitudes generated by distinct input sources to single neurons in the network: external input stimulus v.s. synaptic interactions from other neurons. This ratio always remains below 0.7, indicating that recurrent network connections play an important role in shaping spiking activity in all regimes considered.  See
also~\cite{Lajoie:2013p17297} for further
investigation of this mechanism using spike-triggered
averages.

Even if the stimulus input does not overwhelm the network, its
statistics play an important role in shaping the level of
discriminable features carried by network spike patterns. As
we vary input parameters, two key quantities
change. One is the ``signal strength": as $\e$ becomes
smaller, the direct impact of stimulus fluctuations on neurons subsides, and, as a consequence, so does the magnitude of the differences between stimuli 
$I^A(t)$ and $I^B(t)$. The second is the ``noise": the
chaotic variability in the responses to a single input
signal.  These two
quantities are not independent, and their relationship is
not {\it a priori} obvious. Together, they combine to
create the chaos-induced noise floor described above.  For stimulus parameters that fall below this noise floor, chaotic variability is too
widespread to allow different stimuli to be accurately classified. 

  As described earlier, the variability across spike patterns from the same response ensemble depends on the dimension of the underlying chaotic attractor. There are many ways to quantify this dimensionality, here we use the number of positive Lyapunov exponents divided by the dimension of the system (see {\it Methods} or~\cite{Lajoie:2013p17297} for details of their computation): 
  $$\L^+ \equiv \#\{\l_i>0\}/N.$$
 Intuitively, $\L^+$ indicates the fraction of unstable directions in state space at any given time. As the geometric properties of our network attractors impose that those directions generally align with neural coordinates $\o_i$ (see earlier text about spike-time reliability), $\L^+$ dictates how many neurons are ``unreliable" in the system at any given time~\cite{Lajoie:2013p17297,Lajoie:2014p18333}.
Figure~\ref{fig:parametric} (D) shows $\L^+$ as a function
of $x$. Notice that the network becomes more chaotic
(more positive Lyapunov exponents) as the fluctuation amplitude $\e$ of
the inputs $I_i(t)$ shrinks. This can be interpreted as weaker stimulus fluctuations giving less entrainment of neural
dynamics by the inputs, and hence allowing intrinsic dynamics, the mechanism by which chaos emerges, to dominate (c.f.~\cite{Rajan:2010p7924}). 
 Thus, as $x$ increases, both signal and noise factors should
  conspire to make input stimuli less
  discriminable.  We next quantify this effect, and study how it depends on input similarity (quantified by $\r_{same}$) and readout dimension $N_r$.
  
We numerically estimate the Tempotron's discrimination performance $P$ along the $x$-parametrized curve of stimulus parameters, for a range of $\r_{same}$ and $N_r$ values.  We follow the same procedures described in the previous section. This determines regions of parameter space ($x$, $N_r$, $\r_{same}$) where perfect classification performance, $P=1$, is achieved. The boundaries of these regions are shown in Figure~\ref{fig:parametric} (E), where every parameter point to the left of the boundaries yields perfect classification.  These boundaries have the following interpretation: for a given input parameters specified by $x$, one needs to read-out from a number of neurons $N_r$ greater than a given $\r_{same}$-boundary to be achieve perfect discrimination of inputs with a similarity level given by $\r_{same}$.  As expected, the regions of perfect performance shrink as inputs become more similar. This means that more readout dimensions $N_r$ are needed to discriminate more-similar inputs.  Importantly, the boundaries are positioned at much lower $N_r$ than the corresponding $N_{same}=\r_{same}N$ for all cases, indicating that networks across a broad parameter range can classify inputs using neurons that themselves do not receive discriminable inputs (i.e. $I^A_i(t)=I^B_i(t)$) as demonstrated for our benchmark parameter set in the previous section.
  
Moreover, there is a critical region of stimulus statistics ($x\sim0.5$) where all classification boundaries aggregate for high
$N_r$. This represents the chaotic noise floor for $x$,
beyond which inputs 
cannot be perfectly
discriminated. Figure~\ref{fig:parametric} (F) illustrates this noise floor in more detail, by showing
contour plots of classification performance $P$ in
($x, \; N_r$)-space for three values of $\r_{same}$. It
shows that while $P$ eventually drops to chance (0.5)
as $x\to1$ --- an expected behavior since 
$I^A(t)$ and $I^B(t)$ become indistinguishable when
$\e=0$ --- this transition becomes sharper for higher
input similarity $\r_{same}$. This suggests that when
two stimuli have many identical components (high $\rho_{same}$), the network can either classify
them very well or not at all, depending on the stimulus amplitude. When inputs are significantly dissimilar,
this transition is more gradual. Importantly, for
large enough readout size $N_r$, the noise floor is
almost identical for all input similarities,
indicating that for this network, perfect
classification becomes impossible for any input
similarity once $\L^+$ reaches about $11\%$ of $N$.  

\smallskip
In sum, for the chaotic networks at hand, {\it perfect stimulus 
  classification can be achieved for a
  wide range of stimulus statistics and similarities, and classification can be achieved using spikes from relatively few
    readout cells. However, once the stimulus amplitude falls below a chaos-induced noise floor, classification performance degrades rapidly.}

\bigskip 

%---------------------------------------------------------------------------------------------------
\section{Discussion}

%-----
\subsection{Summary}

Sparse, strongly connected recurrent neural networks used to model cortical activity in the brain often produce a {\it balanced state}, leading to chaotic dynamics~\cite{Vreeswijk:1998p14451}.  
We studied how this chaos -- viewed as an intrinsic source of variability -- impacts the capacity of recurrent networks to accurately encode temporal stimuli.  
With detailed numerical simulations grounded in the theoretical literature, we studied how similar stimuli ---modelled by multi-dimensional frozen white noise inputs--- can be decoded from chaotic network responses, both at the level of the network state space and output spike trains. 

Two factors influence the ability of a decoder to
successfully classify stimuli based on network outputs.  The first is the strength of the chaos-induced ``noise":  
the trial-to-trial variability of evoked patterns due
to chaos.  The second is the ``signal":  the sensitivity of evoked patterns to the
choice of input. 
 Our analysis of these
separate factors leads to three main points: (1)
Chaos in recurrent spiking networks does not, in and
of itself, preclude the accurate encoding of temporal stimuli; simple decoders read out these stimuli based on reliable multi-spike patterns that chaotic networks produce via low-dimensional attractors. (2) Recurrent connectivity distributes stimulus information throughout
chaotic networks, enabling high-dimensional stimuli to be classified with
low-dimensional readouts. (3) Stimulus statistics (i.e., the amplitude of
stimulus fluctuations relative to their mean) modulate
the number of readout neurons necessary to successfully
classify them: as their amplitude decreases, more neurons are required to discriminate stimuli, until a "noise floor" is reached where
discrimination is no longer possible.

 %-----
\subsection{Biological implications of encoding and computing in chaotic networks} 

Chaotic dynamics appear as an emergent property of
recurrent connectivity between neurons~\cite{Vreeswijk:1998p14451,Banerjee:2006p16587,Banerjee:2008p6505,London:2010p10818,Monteforte:2010p11768,Rajan:2010p7924} that would be
otherwise very stable and reliable
(see~\cite{Mainen:1995uz,Bryant:1976ub} and~\cite{Ritt:2003p145,Lindner:2003p16585} for
reliability of single neurons).  
It is conceivable that such chaos is a 
significant contributor to experimentally
observed variability as well
(e.g.~\cite{deRuytervanSteveninck:1997p16808,
  Kara:2000p12837,
  Reinagel:2000p11391,Tiesinga:2008p12846}).  In this way, chaos amplifies and adds to other stochastic noise sources in biological networks.

Beyond contributing variability and noise, there is a substantial literature addressing the
potential advantages of chaotic dynamics for encoding
and
memory~\cite{Buonomano:2009p17775,Sussillo:2030p4602,Laje:2013p17820}. In
many cases, chaotic networks act as ``reservoirs" and
synaptic connections are trained to use their activity
to perform a given task.  
Here, we take a more pragmatic perspective, studying how chaotic networks work as ``channels" that receive
inputs and produce spike outputs that carry usable
information. Furthermore, we show that same recurrent connectivity that produces chaos also serves to
  distribute stimulus information throughout the network:  discriminability is maintained even if a decoder
  only has access to small subpopulations of neurons, and even when the inputs to be
  discriminated do not directly drive the
  subpopulations.  As such,
  recurrence may serve to simplify the process of reading out stimuli from large populations, eliminating the need for precise wiring -- with the resulting chaotic dynamics being a manageable
  by-product.  
  While this type of stimulus ``spread" can also occur in multilayer feed-forward networks with fan-out between layers, a recurrent architecture does the same operation locally, without requiring that decoders be located downstream.  We speculate that this mechanism may also be relevant for contextual
  coding~\cite{Schwartz:2007p1118,Mante:2013p328},
  where the response of some neurons to a fixed
  local input changes if a secondary contextual input
  to others differs.
  This role for recurrence complements many other functions that
  it may serve in neuronal
  computation (e.g., maintaining working memory,
  enabling winner-take-all computation, sharpening
  tuning curves, etc.).

\medskip
Finally, we argue that the encoding mechanism based
on spike patterns we outline in this paper enhances
earlier balanced network encoding mechanisms. 
Classic
  results point to important properties of balanced
  population-averaged activity: its response to global
  external inputs is both rapid --- much faster than
  single neuron time constants --- and
  linear~\cite{Sompolinsky:1988p534}.
If the
inputs to our network evoke different population
firing rates, then population averages carry the
necessary information for discrimination. In contrast,
when two inputs have similar statistics and differ
only in the fine temporal patterns they carry, we show
that the same network can rely on spike-time based
mechanisms to classify them.
It is unclear if such dynamics are present in cortical
circuits and if so, in which regime they typically
operate. However, there is evidence of different
activity states a given cortical network can take
(e.g. up and down states) depending on various
contextual factors~\cite{Mante:2013cn}. In light of the
results we outlined, it is conceivable that cortical
networks encode different aspect of inputs depending
on these input's features. Under this assumption, the
emergent nature of chaos in recurrent networks may act
as a natural mechanism to implement adaptive coding
schemes, without any changes required to the network
or neurons themselves.

%----
\subsection{Future work}

The results presented in this manuscript address a specific class of models, albeit one that is fairly prototypical. Further studies should focus on the effect of single neuron
dynamics and connectivity statistics on stimulus encoding. Moreover, beyond the amplitude effects studied here, the correlation of stimulus inputs across neurons can also impact the resulting chaotic network responses (data not shown).  At the same time, these input correlations effectively diminish the
dimensionality of the stimulus by introducing
redundancies. An interesting area of future work is to
better understand the relationship between input and
output dimension with respect to stimulus coding in recurrent, spiking networks.
 
In experiments, is not an easy task to test
whether or not a particular neural circuit is
chaotic. Indeed, even for a dynamical system that does
not receive input drive, and for which one can
observe all degrees of freedom, it is still a hard
problem to attribute variability to stochastic or
deterministic (chaos) mechanisms (see
e.g.~\cite{Brock:1986wr,Sugihara:1990io}). Therefore,
the problem of experimentally verifying the nature of
variability in neural circuits found in the brain is
not a simple one. Nevertheless, we note that some {\it in vivo}
experiments show stimulus-evoked spikes that appear to have the type of intermittent variability we
described in this article~\cite{Kara:2000p12837,
  Reinagel:2000p11391}. This invites future work to make closer connections between mechanistic models of chaotic dynamics and neural recordings. 

%--------------------------------------------------------------------------------------------------------------------------------------
\section{Methods}
%----------------------
\subsection{Model equations}

The dynamics of recurrently coupled QIF
units follow the formalism of~\cite{Monteforte:2010p11768,
  Lajoie:2013p17297,Lajoie:2014p18333}.  The internal dynamics of a single neuron are given by
\begin{equation}
  \dot{v} = \frac{1}{\tau_{Q}}\frac{(v-v_R)(v-v_T)}{\Delta v}+I_{ext}(t)
  \label{qif}
\end{equation}
where $\tau_Q$ is the membrane time-constant, $v_R$ and $v_T$ are rest and threshold potentials, $\Delta v=v_T-v_R$ and $I_{ext}(t)$ is an external input stimulus. Typical parameter values are $v_R=-65 $ mV, $v_T=-55$ mV and $\tau_Q=10$ ms. The dynamics of~\eqref{qif} are hybrid: once the membrane potential $v$ exceeds the threshold, it blows up to infinity in finite time at which point a spike is said to be emitted and $v$ is manually reset to $-\infty$.
By a smooth change of
  coordinates~\cite{Ermentrout:1996p10447}, one can
  map the voltage interval $(-\infty,\infty)$ onto the
  interval $[0,1]$ in a one-to-one fashion, mapping
  voltages to ``phase'' variables $0\leq\o_i\leq1,$
  where $\o=0$ and $\o=1$ are identified (i.e. $[0,1]$ represents the unit circle).
  The phase of a cell represents the fraction of the
  ``spike cycle'' it has completed; we think of the
  neuron as emitting a spike each time $\o=1$.
  Similarly, the state space of the network is
  the cartesian product of $N$ copies of the unit
  circle, i.e., the $N$-dimensional torus, which one can
 view as an $N$-dimensional cube with opposite faces
  identified.  As formulated in
  Equation~\eqref{net_model} below, this model has
  dimensionless units.  For the sake of clarity, we
  report spikes and other temporal observables using
  milliseconds by fixing the neural time constant to
  $10$ ms in the QIF coordinates (see appendix of~\cite{Lajoie:2013p17297} for more details about this coordinates change).
  
  We consider a network of $N$ neurons separated into
excitatory and inhibitory populations, coupled randomly according to a
Erd\"os-Renyi connectivity with mean in-degree $K \ll N$
from each E/I population. The state of the network at
time $t$ is represented by the vector-valued solution
$\o(t)=(\o_1(t),...,\o_N(t))$ which lives on the
$N$-dimensional unit torus $\T^N$.  The dynamics of
neuron $i$ are governed by \beq
\label{net_model}
\dot{\o}_i=F(\o_i)+Z(\o_i)\sum_{j=1}^Na_{ij}g(\o_j)+\frac{\e^2}{2}Z(\o_i)Z'(\o_i)
+Z(\o_i)\underbrace{[\n +\e \dot{W}_{i,t}]}_{I_i(t)}
\eeq
where $F(\o_i)=1+\cos(2\pi\o_i)$, the phase
    response curve~\cite{Ermentrout:2010p10442} $Z(\o_i)$ is given by
  $1-\cos(2\pi\o_i)$, and $g(\o_j)$ is a sharp ``bump"
  function, nonzero only near the spiking phase
  $\o_j=1\sim0$, modelling the rapid rise and fall of post-synaptic currents
  (see~\cite{Lajoie:2013p17297}). 
  
  The $I_i(t)$'s
  represent external inputs to neuron $i$; here, we
  model these by the sum of a DC current $\n$ and
  independent white-noise processes $\dot{W}_{i,t}$,
  and the $O(\e^2)$ term results from the change of
  variables (we alway interpret white noise forcing in
  the sense of It\^o
  calculus)~\cite{Lajoie:2013p17297}.  Together, the
input signals to each neuron form the global input to
the network : $I(t)=(I_1(t),...,I_N(t))$ as depicted
in Figure~\ref{fig:intro} (A).  Throughout most of the paper, signal parameters are set to $\e=0.5$, $\n=-0.5$ while
$N=500$, $K=20$ and $|a_{ij}|\simeq0.2$ so that the
network is in a fluctuation-driven excitable regime,
producing sustained irregular activity characterized
by a broad firing rate distribution with a mean of
about 6.5 Hz~\cite{Lajoie:2013p17297}. Note that many
of the results presented scale linearly with $N$
when all parameters remain fixed
(cf.~\cite{Lajoie:2014p18333}). We consider
independent inputs $I_i(t)$ across neurons $i=1,...,N$
but briefly address the implication of such correlations in the Discussion
section.

We stress that the $I_i(t)$ model inputs to the
  system, and not the various molecular and cellular
  sources of noise associated with neuronal dynamics.
  In this framework, the response of the network to a
  specific input can be modelled by ``freezing,'' or
  choosing specific realizations of the stochastic
  processes $\dot{W}_{i,t}$~; this is sometimes known
  as a ``frozen noise"
  experiment~\cite{Bry+76,Mainen:1995p16589}.  Given
  specific realizations of inputs, the
  model~\eqref{net_model} can be viewed as a
  deterministic, nonautonomous system of ordinary
  differential equations.  We study
  responses of the network to fixed signals across
  repeated trials where the only source of variability
  arises from changes in the initial condition of the
  network.

%-----
\subsection{Numerical simulations}
A standard Euler-Marayuma~\cite{Asmussen2007} scheme
was used to numerically integrate
Equation~\eqref{net_model}, treated as a stochastic
differential equation. Pseudo-random increments used
to sample the fixed white-noise realizations $I_i(t)$
were generated using the Mersenne Twister
algorithm. A time-step of $\Delta t=0.5$ ms was used for all simulations. We verified that smaller temporal resolution did not change our results. For estimates involving sampling of many trajectories within response-ensembles, initial states of the network were uniformly sampled over state space $\T^N$.  Numerical estimates of Lyapunov exponents were obtained by evolving the adjunct variational equation of~\eqref{net_model}; further details can be found in~\cite{ Lajoie:2013p17297,Lajoie:2014p18333}. Large batched simulations were carried out on the NSF XSEDE {\it Science Gateway} supercomputing platform.

Numerical simulations were implemented in Python and
Cython programming languages. Computations of statistical quantities such as pairwise trajectory distances and spike-time reliability, as well as the Tempotron classifier training and testing, were implemented in MATLAB.

%----------------------------------------------------------------
 \section{Acknowledgments}
The authors thank Fred Wolf and Lai-Sang Young for
very helpful discussions.  GL was supported by a Bernstein Fellowship from the Bernstein Center for Computational Neuroscience, by a postdoctoral fellowship from the {\it Fonds de Recherche du Qu\'ebec} and by an {\it Innovation Fellowship} from the Washington Research Foundation.  KL was supported in
  part by NSF grant DMS-1418775. JPT is supported by a Discovery grant from the Natural Sciences and Engineering Council of Canada (NSERC Grant No. 210977 and No. 210989), operating funds from the Canadian Institutes of Health Research (CIHR Grant No. 6105509), and the University of Ottawa Brain and Mind Institute (uOBMI). ESB and GL by NSF CAREER Grant DMS - 1056125 and NIH Training grant 5T90DA032436.

%=================================================================
%Bibliography
% \begin{small}
% \bibliographystyle{plain}
% \bibliography{ma_bibli_2}
% \end{small}

%Bibliography data

 %-----------
\end{document}